\documentclass[aps,prl,twocolumn,superscriptaddress,amsmath,amssymb,showpacs,showkeys]{revtex4}
\usepackage{epsfig}
\usepackage{prettyref}
\begin {document}
\title {Paramagnetic to Superparamagnetic Transition in Ni(OH)$_2$ Nanoparticles}
\author{S. D. Tiwari}
\affiliation{School of Physics and Materials Science, Thapar University, Patiala 147004, India}
\author{K. P. Rajeev}
\affiliation{Department of Physics, Indian Institute of Technology, Kanpur 208016, India}
\begin{abstract}
We report the temperature and field dependence of $dc$ magnetization on  sol gel prepared nanoparticles of Ni(OH)$_2$. At higher temperature the system is found to behave as a paramagnet while we find evidence for superparamagnetic blocking at low temperature. The system shows a  paramagnet-superparamagnet transition and we discuss the underlying mechanism. 
\end{abstract}
\maketitle
For the past decade or so nanomaterials have been attracting a great deal of attention from scientists and engineers. The increasing interest of scientists, in particular, can be attributed to the novel and unusual properties that the nanomaterials exhibit compared to traditional materials. Nanoparticles are a particular class of nanomaterials and  nanoparticles of magnetic materials have been engaging our attention for sometime now. Nanoparticles of ferro and ferrimagnetic materials are rather well studied compared to nanoparticles of antiferromagnetic materials. In recent years nanoparticles of antiferromagnetic materials have been reported to show unusual and interesting results  never observed in ferro and ferrimagnetic nanoparticles \cite {Kodama 1997,Morup 2004}. Nanoparticles of antiferromagnetic NiO have been rather well studied \cite {Kodama 1997, Richardson and Milligan, Richardson 1991, Makhlouf 1997} compared to other antiferromagnetic nanoparticle systems. This nanoparticle system is known to show many anomalous properties \cite {Kodama 1997, Makhlouf 1997}. Recently we showed \cite {sdt-1, sdt-2} that this system shows spin glass behavior at lower temperatures and this result has been independently arrived at by others as well \cite {Winkler}. In all these works the nanoparticles of NiO of different sizes are prepared by almost similar methods. To begin with Ni(OH)$_2$ is prepared by a sol gel method and then nanoparticles of NiO of different sizes are produced by heating the Ni(OH)$_2$ at different temperatures. Although NiO nanoparticles are rather well studied,  the behavior of Ni(OH)$_2$ itself has not attracted much attention, even though it happens to be quite an interesting nanoparticle system in its own right. Long back, Richardson and Milligan \cite {Richardson and Milligan} reported some work on sol gel prepared Ni(OH)$_2$ where they claimed that the the system is paramagnetic and its susceptibility can be described by Curie-Weiss law at higher temperatures. Interestingly they mentioned that the susceptibility of the system is field dependent below 100~K but this data was not reported and neither was there any follow up work to understand the system. This state of affairs have motivated us to have a look at the sol gel prepared Ni(OH)$_2$ afresh.

Nanoparticles of Ni(OH)$_2$ are prepared, by a sol gel method, by reacting aqueous solutions of nickel nitrate and sodium hydroxide at room temperature at pH = 12, as described in our recent works \cite {sdt-1, sdt-2}. The temperature and magnetic field dependence of the magnetization of this sample is measured with a SQUID magnetometer (Quantum Design, MPMS XL5).

We characterized our sample by x-ray diffraction and transmission electron microscopy. The X-ray diffraction pattern indicated that the prepared sample is single phase hexagonal Ni(OH)$_2$. The average crystallite size was calculated by x-ray diffraction line broadening using the modified Scherrer formula \cite{sdt-1, sdt-2, Scherrer} and it turns out to be about 8~nm. The average particle size estimated from transmission electron micrograph is  close to the  average crystallite size obtained from x-ray diffraction.
\begin{figure}[bt]
\begin{center}
\includegraphics[angle=0,width=1.\columnwidth]{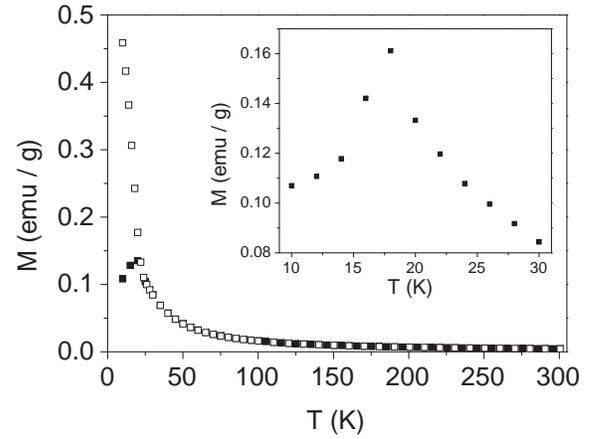}
\caption{ZFC (solid symbol) and FC (open symbol) magnetization for Ni(OH)$_2$ nanoparticles as a function of temperature in a field of 100~G. The inset shows a magnified view of the peak in the ZFC magnetization as a function of temperature.}
\label{fig:mt-dc}
\end{center}
\end{figure}
\begin{figure}[b]
\begin{center}
\includegraphics[angle=0,width=1.\columnwidth]{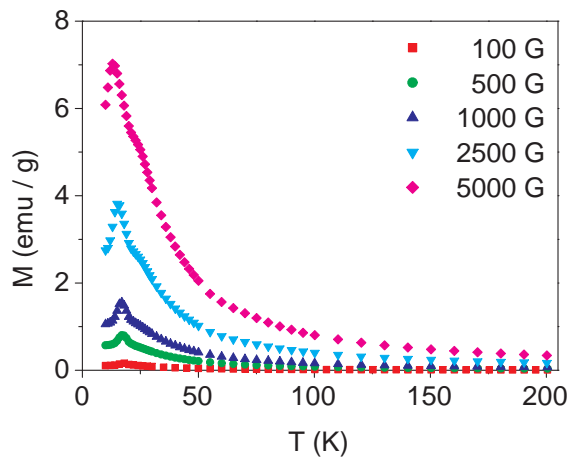}
\caption{ZFC magnetization as a function of temperature for Ni(OH)$_2$ nanoparticles in various applied magnetic fields.}
\label{fig:mt-dc-field-dependence}
\end{center}
\end{figure}
\begin{figure}[tb]
\begin{center}
\includegraphics[angle=0,width=1.\columnwidth]{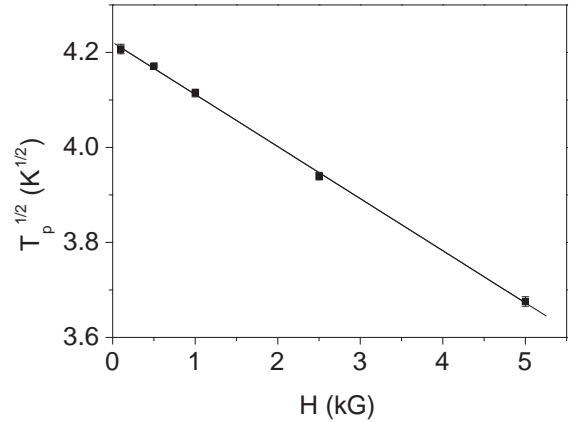}
\caption{Plot of $T_{p}^{1/2}$ as a function of applied magnetic field $H$. The solid line shows a linear fit to the data. The coefficient of determination $R^{2}$ for this fit is 0.99944.}
\label{fig:root-T_p-H-linear-fit}
\end{center}
\end{figure}

We measured the zero field cooled (ZFC) and field cooled (FC) magnetization as a function  of temperature for Ni(OH)$_2$ nanoparticles and the results, for a measuring  field of 100~G, are shown in Figure \ref{fig:mt-dc}. This figure shows that as a function of temperature there is a peak in ZFC magnetization while the FC magnetization decreases monotonically with increasing temperature. These features are characteristic of a superparamagnet \cite{Bitoh}. In the case of a superparamagnet the temperature corresponding to the peak, $T_p$, in the ZFC magnetization is called the blocking temperature. The blocking temperature decreases with increasing applied magnetic field and the field dependence is described by \cite{Bitoh}
\begin{equation}
\label{eq:field-dependence-T_p}
T_p(H) \propto (1 - \frac{H}{H_K})^2,
\end{equation}
where $H_K$ is a constant.  From this equation it is clear that for a superparamagnet the square root of the blocking temperature should decrease linearly with increasing strength of applied magnetic field. To check whether this is the case we measured the ZFC magnetization as a function of temperature in different applied magnetic fields and the curves are shown in Figure \ref {fig:mt-dc-field-dependence}. In figure \ref {fig:root-T_p-H-linear-fit} we plot $T_p^{1/2}$ against applied magnetic field $H$ and note that these quantities are linearly related. The solid line shows a straight line fit to the data which is seen to be very good. These observations indicate that the peak in the ZFC magnetization seen in figure \ref {fig:mt-dc} is due to superparamagnetic blocking of magnetic moments of Ni(OH)$_2$ nanoparticles.

\begin{figure}[tb]
\begin{center}
\includegraphics[angle=0,width=1.\columnwidth]{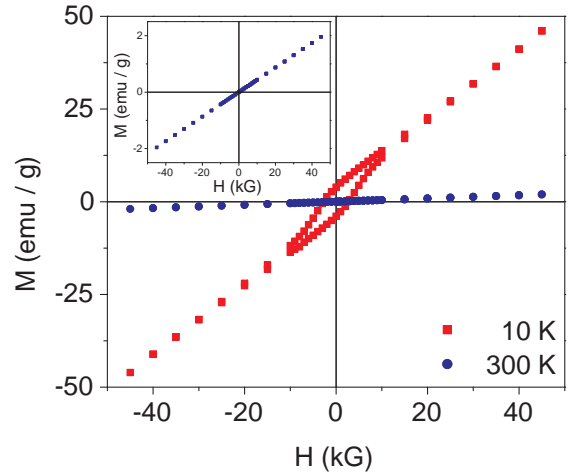}
\caption{(color online) M-H curves at 10~K and 300~K for the sample.  Hysteresis is seen in the 10~K data while there is no hysteresis at 300~K. The 300~K data is again shown in the inset with a different scale to clarify that there is no hysteresis.}
\label{fig:mh-loop}
\end{center}
\end{figure}

We measured the magnetization as a function of magnetic field at different temperatures. This is shown in Figure \ref {fig:mh-loop} at two different temperatures of 10~K and 300~K. We see a hysteresis loop at 10~K, well below the blocking temperature. But at 300~K, well above the blocking temperature, there is no coercive force and no hysteresis.
\begin{figure}[b]
\begin{center}
\includegraphics[angle=0,width=1.\columnwidth]{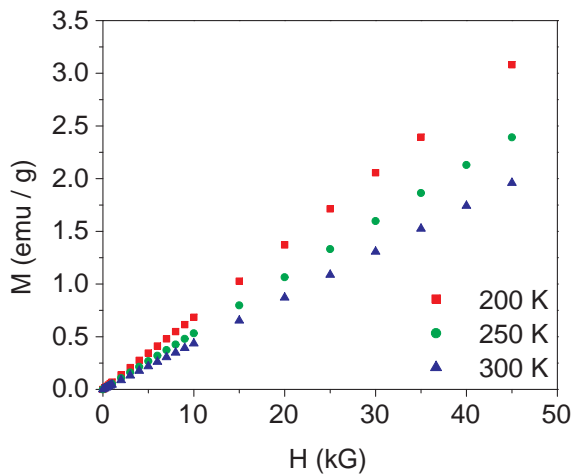}
\caption{(color online) Magnetization $M$ as a function of applied magnetic field $H$ at different temperatures.}
\label{fig:mh-high-temperature}
\end{center}
\end{figure}

\begin{figure}[tb]
\begin{center}
\includegraphics[angle=0,width=1.\columnwidth]{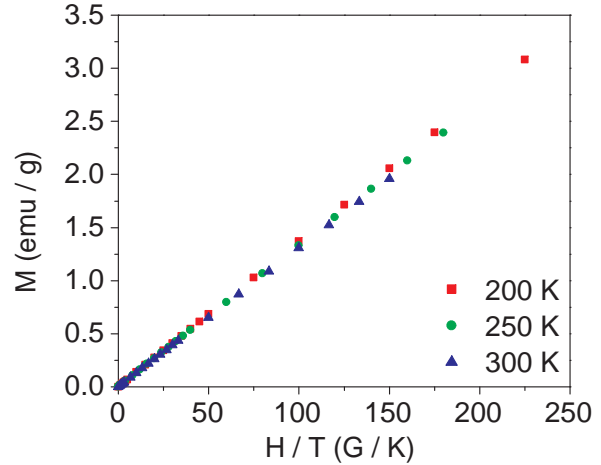}
\caption{(color online) Magnetization $M$ as a function of $\frac{H}{T}$ for Ni(OH)$_2$ nanoparticles.}
\label{fig:mh-overlapping}
\end{center}
\end{figure}

The magnetization of a system consisting $N$ noninteracting magnetic moments each of magnitude $\mu$ as a function of temperature $T$ and magnetic field $H$ is given by the relation \cite{Bean}
\begin{equation}
\label{eq:lang-short}
M = N \mu L(x).
\end{equation}
Here $L(x)$ is the Langevin function defined as
\begin{equation}
\label{eq:lang}
L(x) = [\coth x - \frac{1}{x}],
\end{equation}
where $x = \frac{\mu H}{k_B T}$ and $k_B$ is the Boltzmann constant. Equation (\ref {eq:lang-short}) describes the magnetization of a superparamagnet as well as of a paramagnet. In the case of a superparamagnet $\mu$ denotes particle magnetic moment, usually of a magnitude of a few hundred to a few thousand Bohr magnetons, whereas for the case of a paramagnet $\mu$ denotes the ionic magnetic moment of magnitude of only a few Bohr magnetons. This huge disparity between the $\mu$ values of the superparamagnet and the plain paramagnet shows up as distinct saturation behaviors: the magnetization of a superparamagnet tends to saturate at relatively low applied magnetic fields while a paramagnet requires a huge magnetic field to saturate its magnetization. We have collected the field dependent magnetization data at 200~K and 250~K in addition to the 300~K data shown earlier to check whether our sample shows superparamagnetism at these temperatures. We note that these temperatures are well above the blocking temperature of the system. The data are shown in Figure \ref {fig:mh-high-temperature} and we observe that the magnetization of the system increases with increasing magnetic field or with decreasing temperature as would be expected for a superparamagnet above its blocking temperature. From Equations (\ref {eq:lang-short}) and (\ref {eq:lang}) it is clear that for a paramagnet as well as for a superparamagnet the magnetization $M$ is a function of $\frac{H}{T}$. In other words all $M$ vs. $\frac{H}{T}$ curves should superimpose \cite{Bean}. A plot of magnetization as a function of $\frac{H}{T}$ is shown in Figure \ref {fig:mh-overlapping}. This figure shows that all the data points lie, more or less, on a single master curve. We also see that there is no sign of saturation of the magnetization even at a value of 225~G/K of $\frac{H}{T}$. This plot shows that the behavior of the system is paramagnetic and not superparamagnetic for the following reasons. Firstly, for superparamagnets the plot is not so good because there is always some particle size distribution \cite {Seehra 2000}. Any distribution in $\mu$ will always prevent the data points from lying on a single master curve whenever $M$ is plotted against $\frac{H}{T}$. Secondly, a superparamagnet saturates at comparatively lower values of $\frac{H}{T}$ \cite {Bean, Seehra 2000, Makhlouf} whereas a paramagnet saturates at much higher values of $\frac{H}{T}$ \cite {Kittel, Saturation-Note}. Thus we see that at high temperatures the behavior of the system is paramagnetic.

From the previous discussions it is clear that the behavior of the system is superparamagnetic at lower temperatures whereas it is paramagnetic at higher temperatures. The magnetic susceptibility $\chi$, defined as the ratio $\frac{M}{H}$ at low fields, of a paramagnet decreases with increasing temperature $T$ following Curie law $\chi = \frac{C}{T}$. Here $C$ is a material dependent constant. In paramagnetic materials the interaction energy of any spin with the other spins is zero. A non zero interaction among the spins may result in ferromagnetic or antiferromagnetic ordering of spins below a transition temperature $T_C$. In such cases the susceptibility of the system above $T_C$ is described by Curie-Weiss law $\chi = \frac{C}{T - T_C}$, if the interaction results in ferromagnetic ordering, and is described by $\chi = \frac{C}{T + T_C}$, if the interaction results in antiferromagnetic ordering. We fitted the ZFC magnetization data above 100~K shown in Figure \ref {fig:mt-dc} to these equations and found that the best fit is given by the Curie-Weiss law. This indicates that there is some kind of ferromagnetic interaction among the ionic moments. The values of best fit parameters $C$ and $T_C$ are 1.20 $\times$ 10$^{-2}$ emu K/g Oe and 26~K respectively.

Let us do a quick check to see whether the numbers obtained from the fit are reasonable. According to the Hund rules \cite {Kittel} the values of spin quantum number $S$ and total angular momentum quantum number $J$ for Ni$^{2+}$ are 1 and 4 respectively. This gives rise to a spin angular momentum of 2.83~$\mu_B$ and total angular momentum of 5.59~$\mu_B$. The net magnetic moment of the Ni$^{2+}$ in Ni(OH)$_2$ will depend on the extent of quenching of the orbital angular momentum and will have a value between 2.83~$\mu_B$ and 5.59~$\mu_B$. The value of fit parameter $C$ in Curie-Weiss law is $\frac{N \mu^2}{3 k_B}$. Using the numerical value from the previous paragraph, the value of $\mu$ turns out to be about 2.98~$\mu_B$ which falls in between 2.83~$\mu_B$ and 5.59~$\mu_B$. The number is quite consistent and as an aside we note that the orbital contribution seems more or less completely quenched.

\begin{figure}[tb]
\begin{center}
\includegraphics[angle=0,width=1.\columnwidth]{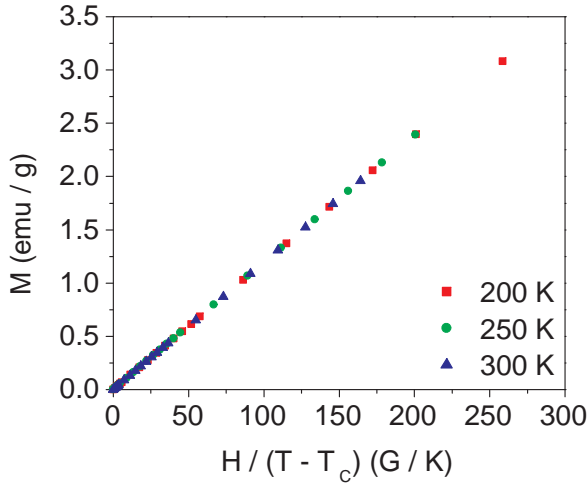}
\caption{(color online) Magnetization $M$ as a function of $\frac{H}{T-T_C}$ for Ni(OH)$_2$ nanoparticles. The extent of overlapping of the data points compared to that shown in Figure \ref {fig:mh-overlapping} is much better.}
\label{fig:mh-overlapping-curie-weiss}
\end{center}
\end{figure}

We find it difficult to resist describing an interesting observation before we wrap up. We have found that the magnetization obeys the Curie-Weiss law and varies as $\frac{1}{T-T_C}$ for $T \gg T_C$. Taking this observation a bit further we may claim that the magnetization $M$ of the system is a function of $\frac{H}{T-T_C}$, instead of $\frac{H}{T}$. This would imply that all $M$ vs. $\frac{H}{T-T_C}$ curves should superimpose. A plot of magnetization as a function of $\frac{H}{T-T_C}$ is shown in Figure \ref {fig:mh-overlapping-curie-weiss} which clearly shows that all the data lie on a single master curve. A very clear improvement in overlapping of the data points compared to that shown in Figure \ref {fig:mh-overlapping} is apparent.

Let us summarize and conclude this story. We reported magnetization measurements on  sol gel prepared  nanoparticles of Ni(OH)$_2$. We find that at higher temperatures Ni$^{2+}$ spins within each nanoparticle behave as paramagnetic ions. As we lower the temperature the system undergoes a paramagnetic to ferromagnetic transition  at an estimated temperature of around 26~K. We would expect that the particles would now be superparamagnetic and, indeed, that turns out to be the case as we see superparamagnetic blocking below about 18~K in a 100~G applied field. We also note that the blocking temperature varies with the applied field as one would expect for a superparamagnet. The paramagnet-superparamagnet transition we have seen appears to be a new finding.

\end{document}